# Using conceptual blending to describe emergent meaning in wave propagation

Michael C. Wittmann, University of Maine, 5709 Bennett Hall University of Maine, Orono ME 04469-5709, wittmann@umit.maine.edu

**Abstract:** Students in interviews on a wave physics topic give answers through embodied actions which connect their understanding of the physics to other common experiences. When answering a question about wavepulses propagating along a long taut spring, students' gestures help them recruit information about balls thrown the air. I analyze gestural, perceptual, and verbal information gathered using videotaped interviews and classroom interactions. I use conceptual blending to describe how different elements combine to create new, emergent meaning for the students and compare this to a knowledge-in-pieces approach.

## Introduction

In several strands of research into student understanding of specific science content, one research assumption is that ideas (be they large scale (Wandersee, 1994; Carey, 1985) or small scale (diSessa, 1993; Hammer, 2000)) already exist and are triggered in a given context, ready to be used, fully formed. The ideas are given various names within physics education research, either misconceptions such as "impulse theory" (McCloskey, 1983a,b), phenomenological primitives (p-prims) such as "force as mover" (diSessa, 1993), or reasoning resources such as "actuating agency" (Hammer, 1996). Across all these different traditions, the research involves modeling which pre-existing ideas are activated and how they are used with each other in a context.

Ideas are not only activated. New ideas must come into existence, else learning would not occur and our toolbox of useful ideas would never increase. This prompts the need for explanations of the development of ideas, be they thought of as conceptions, p-prims, or resources. One could assume a long-term development of an idea through many processes. A rich literature on the development of misconceptions (see Wandersee, 1994, for a review) suggests that misconceptions are formed through inappropriate teaching, misinterpretation of observations of the world around us, or describing situations using inappropriate models. Primitives, on the other hand, might develop very early in life (Mandler, 1992), suggesting that some kinds of p-prims are nearly hard-wired into our thinking. We might reify the several processes we use to solve a problem into a single new cognitive object (Dreyfus 2002; Sfard, 1991; Tsamir, 2004). We can also think of connecting many different resources into a consistently activated network appropriate for a given setting (Hammer et al., 2004) or we can discuss ways in which networks of resources rearrange so as to describe different types of conceptual change (Wittmann, 2006). The plasticity of such networks can be defined, as well (Sayre, 2008). In summary, one can consider many pathways to the development of robust ideas which are later able to be activated.

Not all ideas need be developed in a robust fashion and for long times. Indeed, it is clear that new ideas must arise in new and novel situations. Such ideas might develop on the time scale of seconds and last for only as long as a problem needs solving. Thus, it is appropriate to ask if we can observe such events occurring: do we see students using an idea on very short time scales, and, if so, what observations allow us to conclude so?

In this paper, I suggest one approach to modeling emergent meaning in students who are solving physics problems. I look at common students' responses to a typical physics question about wave motion on a long, taut spring. Students were asked to describe how one creates a single wavepulse on the spring, and then predict how one might change the speed of the wavepulse. I show how students commonly use similar language and gestures as they give their answers. In summary, their answers are similar to descriptions given to describe how a ball is thrown through the air. To model the details of their answer, I use a certain level of formalism that is borrowed from the field of language development and linguistics.

I use the mechanisms of mental space integration, also called conceptual blending (Fauconnier and Turner, 2002), in particular the ideas of composition, completion, and elaboration. In this analysis, I consider two different mental spaces, one being the observed event of a wavepulse being created by a wrist flick and traveling along a long, taut spring, the other being the imagined space of a ball thrown in the air. These two spaces are connected by similar elements, forming a blend composed of elements of both spaces. New information is recruited from the different mental spaces, but only some information from each space is used. The idea represented in the blend is then elaborated upon, leading to emergent meaning. Specifically, students predict that flicking one's wrist harder leads to a faster wavepulse (as if throwing a ball harder). This response seems obvious within a simple metaphor (a wavepulse is like a ball) but gives us insight into ways in which student responses emerge rather than are activated. In particular, one well-documented idea, the Ohm's p-prim (diSessa, 1983, 1993) can be thought of as emerging from the blend rather than being a pre-existing idea that is activated in a new setting. This result is consistent with observations of students engaged in a time-consuming construction of new meaning as they explain their thinking.

## Student responses to a wave propagation problem

The data analyzed in this paper come from interviews where students make predictions about a basic physics problem that is commonly taught and studied in a high school or college survey physics course. Data presented in this section are taken from the author's 1998 Ph.D. dissertation, and have previously been analyzed in several ways (Wittmann 1999, 2002). Data on the question discussed here were gathered from hundreds of students in written questions (free response, multiple choice, and multiple-choice multiple-response questions on ungraded quizzes and specially designed, ungraded pre- and post-instruction surveys, as well as graded tests), tens of interviews, and hours of informal classroom observations. Rather than analyze all students, I am interested in a class of common responses for which I will give a few brief examples. The examples below illustrate the most typical response given by students. Data come from transcripts from interviews carried out with students who were made aware of the Think Aloud protocol (Someron, 1994). The role of the interviewer was to elicit responses about hypothetical situations and explore the reasoning students gave for their responses. Though the original intent of the interviews was not to explore the development of ideas, the data lend themselves to a re-analysis in terms of emergent meaning arising in new and novel situations.

The physics problem concerns wavepulse propagation along a long, taut spring. In a typical form of the interview question, students were given an image of an already propagating wavepulse (see Figure 1). They were asked to describe how such a wavepulse could be created. Then, they were asked how they might change the amount of time it took for the wavepulse to reach the wall; some students were asked to increase the time, some to decrease it. A correct response to such questions would have been to consider the tension in the spring (changing how tightly the spring is pulled) or changing the mass density of the spring (by replacing the spring with a different one, basically). The hand motion used to create the wave does not affect the speed of propagation. (A complete analysis of this situation is found in typical introductory physics textbooks and was part of instruction that all students had completed by the time of these interviews.)

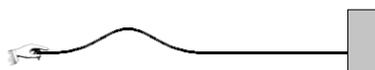

Figure 1. Propagating Wavepulse. A hand moves up and down
and creates a pulse which moves toward a distant wall.

Three examples illustrate common features of many student responses. "Ford" (all names are aliases) had already demonstrated how to create the wavepulse, making a brief up-and-down motion with his hand to show how the shape came to be. The following dialogue occurred shortly after, with information about his drawings and gestures included parenthetically:

IVW: Let's say that you are standing over there, you create it, and it takes a certain amount of time for it to reach the wall. Let's say you want to shorten the amount of time that it takes. What would you have to do?
Ford: One -- I could probably -- there are two scenarios that I have to think about, and since you want me to say right now -- I'd send a quicker one (he draws a pulse smaller than the original in both width and amplitude and makes an up-and-down hand motion in less time and with lower amplitude than the original)
I: Wait, let me stop you right here, by quicker you mean, you did your hand motion like this (copies Ford's small-amplitude motion) --
F: Shorter, I wouldn't go (repeats original large-amplitude hand motion) I'd try to make a shorter hand motion (repeats small-amplitude hand motion), I wouldn't want to like this (repeats original large-amplitude hand motion). It would get there faster.

Notably, Ford's explanation of "I wouldn't go" involved a grammatical pointer to his hand motion, the gesture acting as the predicate to the sentence. This first explanation was immediately followed by a second scenario in which Ford described a larger and therefore faster pulse instead, again with associated hand motion. This dialogue was long and is left out because essential features of the explanation have already been presented – the differences in hand motion and language pointing to them.

"Adam," when asked a similar set of questions, had given the same kind of explanation for wavepulse creation, a flick of the wrist up and down. His explanation used slightly different physics terminology, but involved the same sense of a change to how the wavepulse is created:

IVW: Imagine that we can measure, just clock the time it takes the pulse to go all the way down to that wall. How could you shorten the amount of time that it takes for it to reach that wall, what could you do to make the time less?
Adam: If we could make the initial pulse fast, if you flick it, you flick it faster.
I: ...And why would flicking it faster make it go faster?

A: That would make it, I don't know, it's kind of hard to explain
I: Do the best you can
A: It would put more energy in or something.

Like Ford, Adam emphasized the connection between the hand motion and the speed of the wave. Another student, "David," gave a similar response to Adam and Ford's when he said "I think possibly, you see a slower pulse … if the force applied to the spring is reduced ... that is: the time through which the hand moves up and down [is reduced]." Again, the "force applied to the spring" describes how the hand moves up and down and affects the subsequent speed of the wave.

Many students' explanations were strongly connected to the hand motions they made to indicate the creation of the original wavepulse. The "faster flick" or the "greater effort" or the "smaller wave" were all shown through gesture, typically a harder hand jerk, a quicker flick of the wrist, a more vigorous and robust arm movement, and so on. Their motions and words were consistent with the idea that they were thinking as if the wavepulse were like a ball. Formalizing this intuitive sense of their responses is the purpose of the next section.

## Summary of Conceptual Blending

My goal is to account for the ways in which students used hand and arm gestures and a specific language to arrive at a certain kind of prediction about how to change the speed of the wave (involving changes in effort in creating a wave). It would be facile to claim that they activated the idea that more effort leads to more speed in the face of resistance of the string to being moved (an application of the Ohm's p-prim (diSessa, 1983, 1993) in the context of wavepulse propagation). But, the data suggest that one can think of the idea not being activated (as Ohm's p-prim) but as emerging in this context. To describe emergent ideas, I will use the formalism of mental space integration (commonly called conceptual blending), a framework developed by Fauconnier and Turner (2002) to account for meaning generation in language. Briefly, mental space integration assumes that there are different mental spaces (ideas associated with a given situation) that can combine in specific ways to create new meaning. Distinct mental spaces are combined due to some shared content or structure. The two spaces are brought together ("blended"), with selective projection taking some information from each input to compose a blend. New structure and new information is recruited (perhaps from long term memory) to complete the blend. One can let the blend "run", i.e., let the newly developed idea be elaborated upon. Emergent meaning arises as the results of this elaboration are connected by backwards projection from the blend back into at least one input space, perhaps both. We observe all these elements in student responses to the wave physics question.

I should be clear that I do not consider the analysis of student responses to be a linguistic problem, but instead am using representations and mechanisms from mental space integration to analyze reasoning in physics. I present applications and explanations in the wave physics context below; more detail can be found in books (Fauconnier and Turner, 2002) and in special issues of the journals of Cognitive Linguistics (as summarized in Coulson and Oakley, 2000) and the Journal of Pragmatics (as summarized in Coulson, 2005). The discussion that follows is particularly indebted to Bache (2005) and Hougaard (2005). I introduce the idea of blending with an example that summarizes many of the ideas used later in this paper to describe reasoning about wavepulses. It is adapted from Fauconnier and Turner (2002).

Consider a situation in a lab where you and a careless colleague are setting up an instrument. Your careless colleague is across the room, and about to attach one piece of equipment to another in a way that will cause trouble. You blurt out, "If I were you, I wouldn't do that!" Your colleague stops. What's going on here?

We can imagine two input spaces. You (and what you would do) exist in one space, your colleague in the other. In the blended space, the two of you are fused into a single, unique person ("if I were you" turns into "I am you"). This process of creating a new entity (of you and colleague as one person) is a "composition," in that it is composed of elements from both inputs. This new individual has an agency that is not yours (it is your colleague's) but makes a decision that is not your colleague's (it is, instead, yours). You have knowledge of the system that your colleague does not (which guides your statement "I wouldn't do that!"). This information "completes" the blend. The projection of information from each input space is selective, in that the judgment from your space is used for the decision in your colleague's space. In the blend, your knowledge is applied to you-and-colleague-as-one. You "elaborate" on the knowledge by implying that something should not be done. New meaning emerges – there is danger here. From here, there is back projection, from the blend back into your colleague's mental space: your actual colleague (thankfully) acts differently than originally intended.

To summarize important terms from this example, there are three major aspects that drive blending:
- Composition: creating relations that are not necessarily obvious, but allow the blend to occur
- Completion: recruitment from long term memory, for example in pattern completion or assigning properties to blends based on knowledge from one input
- Elaboration: "running the blend;" imaginative mental simulation that leads to emergent meaning that was not necessarily part of either input

Further terms of importance are selective projection (not all elements of each input space go into the blend, since often there are contradictory elements in the input spaces) and backward projection (information going back from the blended space to at least one input space). The full formalism of mental space integration is much larger than these five elements and includes heuristics for how distinct mental spaces are connected. That formalism is helpful in modeling student reasoning and provides additional details (such as the concept of the compression of cross-space connections), but is outside the scope of this paper. In the next section, I apply the formalism to the data presented above.

## Conceptual Blending of Wavepulses

The formalism of conceptual blending allows us to take student responses (including language and gesture) and make sense of their responses as emergent meaning, without activation of pre-existing ideas. I describe two individual inputs, show how they blend into a single space, and describe the emergent meaning that arises in the blend. I give two examples, the first based on the data that we have of student responses to the interview question, the second a hypothetical discussion which shows that a different set of inputs might lead students to create a blend in which the correct answer emerges. In the following section, I compare this analysis to past analyses (including my own) using another model of student reasoning.

### Seeing waves as balls

One can describe two mental spaces in use when predicting how to change the speed of a wavepulse translating down the spring. First is the simple perceptual observation of the wavepulse, a bump on a spring created by a vigorous hand motion and moving down the spring. Second is one of balls or other finite sized things that can be thrown through space. An understanding of object motion happens extremely early in life (Mandler, 1992) and ideas about thrown balls are readily available to most children (and university students). How these ideas get connected to the wavepulse is described by mental space integration. The result is a blended space with emergent meaning: wavepulses "as" balls, with certain properties that are not there in the observation itself but play out in predictions, when the blend is "run."

A blend is typically described by a diagram such as shown in Figure 2. In the upper left, there is an input related to the observed wave on the spring. As shown by the students, there is a kinesthetic (proprioceptive) element of the wrist flick. One observes a pulse, as well as its motion (translation) down the spring. In the upper right, there is an input that contains typical properties of thrown balls. The representation takes into account issues of selected projection from each input into the blend. The Observed Wave input contains only a selective set of the observed phenomena, while the Ball input contains only a selective set of properties associated with balls. Only those elements of the Ball input which match to elements of the Observed Wave input are shown.

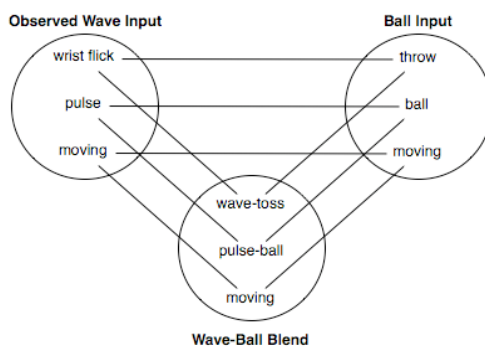

Figure 2. Wave-Ball blend. Two inputs (of the observed wavepulse and of thrown balls) are integrated to create a blend that leads to emergent meaning.

The Observed Wave and Ball input spaces are connected by like elements and integrated to create the Wave-Ball blend. There is selective projection, implied already because of the presentation of only selected pieces of information in each input. The wrist flick (perpendicular to the direction of motion) and a throw (in the direction of motion) are connected because they are both motions that initiate movement. The observation of a finite size pulse is connected to that of finite size objects, perhaps through a perceptual process of *binding*, a psychological term that describes how objects are seen as a whole. The act of binding is impossible for the (healthy) mind to stop. The motion of the pulse down the spring and a thrown ball through the air are connected because they are both propagation in a specific direction. Certain information is ignored: the perpendicular motion of pieces of the spring, for example, and information about the parabolic path of a thrown ball. It is probably unavoidable, based on the creation (by hand flick), shape (finite size), and directed movement of the wavepulse, to *not* think of the wavepulse as a finite-sized object like a ball.

In the blend, one can recruit information from the Ball input, namely that the speed with which you move your hand affects how fast a ball moves through the air. Furthermore, student predictions about how to change the speed of the wavepulse require that they "run" the blend. This leads to emergent meaning that was not present in the Observed Wave input. Notably, the hand motion perpendicular to the direction of wavepulse motion is taken as analogous to the throw of a ball in the direction of motion – obviously, they are not the same, but they are treated as such in the blend. So, in the blended space, the speed with which you move your hand affects the speed of the wavepulse along the spring; students like Ford, Adam, and David either describe a "quicker, faster hand motion" or that you "put more force in your hand" (with an accompanying larger/harder/faster wrist flick) when giving this answer. In the blend, the idea emerges that greater effort leads to greater speed. The Ohm's p-prim is not activated, it emerges.

## Seeing waves as events

One instructional goal of teaching students about waves has been suggested as helping students see the wave as a propagating event rather than an object translating down the spring (Hammer, 2000). For mental space integration to be useful, it must also describe what we believe expert physicists are doing when they think about waves. In Figure 3, we propose a blending diagram to describe how a physicist might model this situation.

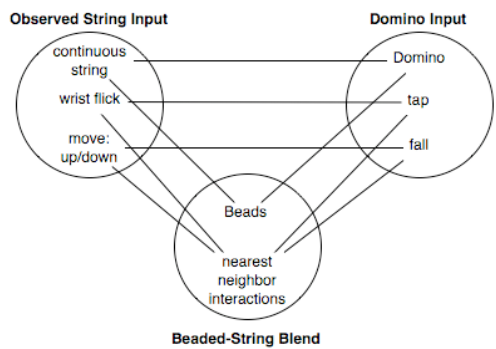

Figure 3. Wave-as-Event blend. In the blended space, the propagation of a wavepulse along a beaded string is an emergent phenomenon.

The blending diagram for this situation again contains two input spaces and a blend. The scale of observation is changed – rather than an Observed Wave (where one focuses on the propagating bump on a long, taut spring), one looks at the spring and has an Observed Spring Input. Three observations can be made. First there, is a long, continuous (taut) spring. Second, there is a flick of the wrist that is holding the end of the spring. Finally, one observes that the spring moves up and down (assuming, as always in this paper, a transverse wavepulse as shown in Figure 1) first at the end, and then consecutively at locations further down the spring. In the upper right of the figure, I suggest a second input: Falling Dominoes. It is important to note that any input containing effects spreading from discrete points to discrete points is sufficient. Examples include news spreading through gossip or a pile-up in a car crash during a traffic jam. The Dominoes example has the benefit that it is part of most people's everyday experience, is easily imagined, and is linear. In the Dominoes input space, there are individual dominoes, lined up in rows. One is tipped over, hitting the next. They fall down sequentially, each being knocked down by the one before it.

The blend is created through several connections. The spring and the Dominoes are both in a long line, so are visually similar. The motion of the wrist up and down at a single point (the end) of the spring is connected to the motion of the first Domino as it is tipped over. Just as each piece of spring goes up and down as it is pulled by the piece of spring next to it, the Dominoes keep falling over and hitting the next Domino in line. We have three elements: the long line, the repeated motions, and propagation of these events along the line.

The blend is composed of elements from each input. In the blend, the continuous spring is re-interpreted as being made up of discrete elements – the structure of the line of Dominoes is applied to the spring. Such a description was already implicitly used in the preceding paragraph, where "each piece of spring" acts on the next (this is not actually true in a continuous system). So, we have a Beaded-Spring in the blend. It moves according to the rules of the Observed Spring, where pieces of the spring go both up and down rather than simply falling over like Dominoes do. The blend, then, is a way of imagining the continuous spring as a set of discrete points that interact with each other through nearest neighbor interactions (like the discrete Dominoes) and move like the observed spring (rather than the discrete Dominoes).

The wavepulse is a global, emergent phenomenon of the blend. It arises because of elaboration, in which one "runs" the blend and plays out the consequences of the wrist flick and nearest neighbor interactions. The up motion of the wrist creates a signal down the spring; nearest neighbor interactions create the leading edge of the wave. The wrist changes direction; the peak of the pulse is created and propagates through the system as each bit of spring recapitulates the motion of the wrist. The signal ends when the wrist returns to its starting position; the wavepulse is completed and propagates through the system. The actual rules of propagation are not observed and only emerge in the blend. The speed of wave propagation depends on the nearest neighbor interactions due to tension and the mass of each individual "piece" of spring. We note that a typical textbook derivation of the wave equation describing wave propagation shows exactly this decomposition of a continuous spring into discrete segments and interactions among the nearest neighbors as the mechanism for propagation

What separates the Beaded-Spring blend from the Wave-Ball blend is the selective attention to the physical system being observed. It should come as no surprise that one reasons differently when attending to different details of a situation. Hougaard (2005) and Bache (2005) describe the choice of what to attend to as a process of disintegration, as one determines which parts of a system to use in the process of selective projection into the blend. In both cases, a motion acting on a system is connected to a different motion acting on a different system. One's choice of system, balls or Dominoes, suggests different ideas of how to make sense of an observation. I do not pursue the analysis further in this paper, except to point out that choosing to observe either the "wave on a spring" or the "spring in motion" allows for different ways for dis-integrating the observation into blendable pieces.

## A knowledge-in-pieces analysis, instead

The analysis given above is not the only analysis possible with the simple interview data presented earlier. In this section, I review a previous analysis of wave propagation on a long, taut spring (Wittmann, 2002). Another possible analysis might be to use elements of the resources framework, a knowledge-in-pieces schema model of reasoning that builds off of diSessa's work (1983, 1988, 1993) on phenomenological primitives and is described more generally by Hammer (1996, 2000, 2004). Many of the ideas of the resources framework are consistent with Marvin Minsky's descriptions of frame systems (1975) and agents (1985). A more general review of the resources framework as used in this section is provided elsewhere (Wittmann, 2006).

## Resources as knowledge pieces

Resources are ideas that are useful and productive when solving some problem (Hammer, 2000). They are basic ideas one has that apply to a situation; they can be thought of as individual, nestable bits of knowledge.

As an example, "closer means stronger" (Hammer, 1996) implies that if you sit closer to a loudspeaker, the sound is louder; if you sit closer to your loved ones, you love them more; if you are closer to the sun, the heat is greater and it's summer. The first is typically true, the second of indeterminate veracity, and the last is false. The resource "closer means stronger" has no inherent rightness or wrongness. It's simply an idea, a knowledge bit, applicable (or not) in a given setting, and useful (or not) when engaged in problem solving. The term "resource" was originally meant to be very general, in the sense of computing, where a resource can be a printer (a tool outside the computer but useful for a program inside the printer), an API (in the software), or a re-usable object (in the code). One thinks of resources as activated or not. Typically, authors have been unclear about the origin of resources (whether they are built on the fly or exist as pre-compiled pieces ready to be activated), suggesting that both may be the case. In this section, for the sake of comparison with the previous discussion of blending diagrams and emergent ideas, we assume that they are pre-compiled and activated.

Few problems are solvable using a single resource; typically, many are necessary. Resources might coordinate with each other, as described by coordination classes (diSessa and Sherin, 1998). Furthermore, resources are useful at different scales - some very complicated concepts can act as resources in a problem solution (e.g., "force" as a primitive in the sense of "force as mover" (diSessa, 1993)) but can also be thought of as being made up of many individual resources (e.g., the coordination of resources and readout strategies about when to use them, described by the coordination class "force" (diSessa, 1998)).

## Resource graphs of wave physics reasoning

I rarely think of students (in interviews or classroom discussions) as thinking in terms of large-scale concepts (Wittmann, 2006). Instead, a simple model of students assumes that they activate only those parts of a concept that are relevant to the situation they are discussing. Similarly, I rarely expect that they use only one resource in a context, since a problem typically requires linking together more than one resource. I refer to the in-between level of description, neither primitive nor concept, as a mesoscopic description and use resource graphs (Wittmann, 2006) to represent this space of student thinking. A resource graph (see Figure 4) is a simple graphical representation of the resources activated by an individual in a particular setting.

Three resources, two previously described in the literature, can be used to account for student responses. First, and simplest of all (but important for later analysis) is Motion, specifically the translation of a bump along a long, taut string. This translation is caused by some startup effect, an Actuating Agency (Hammer, 1996), which diSessa has called Force as Mover (diSessa, 1993). Finally, increased effort in the face of resistance of the spring to being moved leads to greater speed, an example of the Ohm's p-prim (diSessa, 1983, 1993). These three resources are represented in Figure 4.

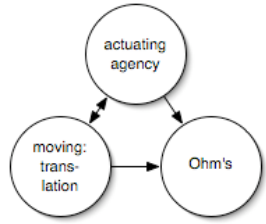

Figure 4. Resource graph to describe the typical response of thinking of wavepulses like balls.

Much as both the common student response and a hypothetical correct response could be modeled through a blending diagram, one can represent a hypothetical correct response using a resource graph, as well. As with the resource graph of waves-as-objects, the resource graph of waves-as-events contains only three elements (see Figure 5). There is an Actuating Agency that is, this time, carried out specifically on the spring. There is a displacement away from and back toward equilibrium for the end of the spring. Through Nearest Neighbor interactions, each element of the spring acts on its neighbor as it was acted upon. Finally, there is Motion, namely the propagation of the up-and-down motion of the hand.

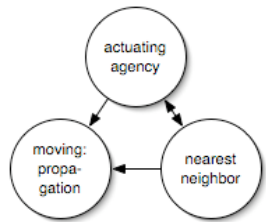

Figure 5. Resource graph for waves seen as events.

## Comparing Blending Diagrams to Resource Graphs

Mental space integration provides a more complete description of student thinking than does the resources framework because it makes fewer assumptions and has greater explanatory power.

There are fewer assumptions made about the nature of pre-existing ideas such as the Actuating Agency resource and how it is applied to different systems. In the blends, the connection between hand motion and the creation of the wavepulse or the motion of the spring is clear. In the resource description (as summarized in the graphs in Figures 4 and 5), one must make more assumptions about how the idea is applied – why was Actuating Agency connected to the wavepulse or to the spring? The use of the resource in its specific context is underdefined within the resources framework, while mental space integration shows directly how an idea is used in a setting.

Mental space integration also gives added explanatory power, indicating how the Ohm's p-prim emerges in a situation that is novel to students (as shown by their seeking many different explanations for their predictions). We need not assume that the Ohm's p-prim already exists and is activated in this setting. Instead, we can describe that the idea emerges, without assumptions of its pre-existence, and can be specific about what aspects of the blend led to its emergence. I do not doubt that there are cases where the Ohm's p-prim is activated as a pre-existing idea, but suggest that in this situation, it emerges. I generally believe that students have a self-checking mechanism ("does this idea make sense?"), and believe that they apply this and find some consistency between the Ohm's p-prim and their emergent idea. There is no evidence of such self-checking, though. It may be that the Ohm's p-prim acts as a "primary metaphor" to guide student thinking about a problem (Grady, 2005) and requires no self-checking mechanism. This suggests a way to combine the resources and blending analyses.

The idea of emergence within a specific context helps clarify some ambiguity in the resource graphs shown in Figures 4 and 5. Motion as a resource was related either to the translation of an object-like wavepulse or the propagation of a wavepulse; the blending description gives greater detail. In the Wave-Ball blend, the wavepulse moves like a ball, and a greater hand motion causes a greater wave speed. In the Beaded-String blend, the wavepulse emerges from elaboration of the blend. These are two very different descriptions.

Finally, modeling student responses with mental space integration provides a predictive power that is not possible in the resources framework. For example, if one throws a rock in a pond, water waves travel outward. An actuating agency caused a motion – but would one predict that a larger rock would cause a faster motion? Yes, the splash is larger; no, the speed is not. The resource graph in Figure 4 does not account for such detail, while the blending diagram in Figure 2 helps us recognize that the two situations might not be connected at all. Without a hand motion, the ball-space ideas simply won't be connected to the water-wave situation and it seems reasonable to predict that students would not think that a larger splash leads to a faster wave in the pond. Similarly, even though a hand is needed create a sinusoidal wave, its motion is not the same wrist-flick as when creating a wavepulse, and the idea of a throw might not exist to connect the one space to the other. Indeed, we find that few students talk about flicking the wrist harder when predicting how to increase the speed of sinusoidal waves.

Modeling student reasoning using mental space integration does have inherent problems. The description is more difficult, including issues of composition, completion, and elaboration (as well as details only alluded to in this paper, such as the compression of cross-space connections). In its favor, the assumptions about students' pre-existing knowledge are fewer, the explanatory power is greater, and the approach allows for a finer grain analysis of what is, and is not, affecting student reasoning in a given context.

## Acknowledgments

The work described in this paper was funded in part by NSF grants REC-0633951 and DUE-9455561. The data were gathered while the author was a graduate research assistant at the University of Maryland.